\RequirePackage{fix-cm}
\documentclass[twocolumn]{svjour3}          
\smartqed  
\usepackage{graphicx}
\usepackage{amssymb}
\usepackage{float}
\usepackage[autostyle=false, style=english]{csquotes}
\MakeOuterQuote{"}
\usepackage{amsmath}
\usepackage[dvipsnames]{xcolor}
\usepackage{verbatim}
\usepackage[hyphens]{url}
\usepackage[hidelinks]{hyperref}
\hypersetup{breaklinks=true}
\usepackage{cite}
\begin{document}


\title{{Different eigenvalue distributions encode the same temporal tasks in recurrent neural networks.}
}


\author{Cecilia Jarne
}


\institute{C. Jarne \at
              Departmento de Ciencia y Tecnolog\'ia de la Universidad Nacional de Quilmes - CONICET \\
              \email{cecilia.jarne@unq.edu.ar}           
}

\date{Received: date / Accepted: date}

\maketitle

\begin{abstract}
Different brain areas, such as the cortex and, more specifically, the prefrontal cortex, show great recurrence in their connections, even in early sensory areas. {Several approaches and methods based on trained networks have been proposed to model and describe these regions. It is essential to understand the dynamics behind the models because they are used to build different hypotheses about the functioning of brain areas and to explain experimental results. The main contribution here is the description of the dynamics through the classification and interpretation carried out with a set of numerical simulations. This study sheds light on the multiplicity of solutions obtained for the same tasks and shows the link between the spectra of linearized trained networks and the dynamics of the counterparts. The patterns in the distribution of the eigenvalues of the recurrent weight matrix were studied and properly related to the dynamics in each task.}

\end{abstract}

\section{Introduction: \label{INTRODUCCION}}

{Recurrent neural networks ({or RNN}) are used to model different areas of the brain such as the cortex. These areas have high recurrency in their connections}, even in early sensory areas that receive stimuli from subcortical areas \cite{Murphy2009}. {Several approaches have been proposed considering RNNs to model these regions, including different network's topologies, connectivity models, architectures and training methods} \cite{SUSSILLO2014156, BARAK20171, NIPS2019_9694}. The advances made in the field have been guided by results obtained in different experiments such as multiple single-unit recording or neuroimaging data \cite{10.1371/journal.pcbi.1005542, Pandarinath2018}.

 {Other models}, as ORGANICs (Oscillatory Recurrent Gated Neural Integrator Circuits) \cite{Heeger22783}, have been inspired by the progress in the field of Machine Learning \cite{Hassabis2017}. In this field, {architectures} such as LSTM (Long Short Term Memory units) and GRU (Gated recurrent units) are widely spread and have been used to process temporal sequences since they do not have the same limitations as RNN to process long time dependencies \cite{69e088c8129341ac89810907fe6b1bfe, NIPS2015_5955, pascanu2013, 279181, 10.1162/neco_a_01253}.

However, the simple RNN model still constitutes a vast field of study. The main reason is that it is used to understand neural computation in terms of collective dynamics, {which} is involved in motor control, temporal tasks, decision making, or working memory \cite{doi:10.1146/annurev-neuro-092619-094115}.

{Understanding the dynamics behind such models allow us} to construct different hypotheses about the functioning of the brain areas and explain the observed experimental results \cite{BARAK20171, KAO2019122}. {For example,} recurrent trained networks have also been recently used to transfer the learned dynamics and constraints to a spiking RNN in a one-to-one manner \cite{Kim22811}.

It has long been known that network's dynamics {is} strongly influenced by the eigenvalues spectrum of the weight matrix that describes synaptic connections, therefore the significance of studying such distribution is to elucidate the different aspects of the dynamic behaviour of the system. 

{For example, the existence of spontaneous activity is related to the synaptic weight matrix and depends on whether the real parts of any of the eigenvalues of its decomposition are large enough to destabilize the silent state in a linear analysis.} Also, the spectrum of eigenvalues with larger real parts provides signs about the nature of the spontaneous activity in full nonlinear models \cite{PhysRevLett.97.188104}.

{The correlation} between eigenvalue spectra and dynamics of neural networks has been previously studied in \cite{doi:10.1162/neco.2009.12-07-671} relating the design of networks with memory face with the eigenvalues outside the unit circle in the complex plane, {which means networks that have attractor states (or memories) according to \cite{doi:10.1162/neco.2009.12-07-671}.}

This spectrum has been the subject of study under different {hypotheses} and connectivity models \cite{PhysRevLett.97.188104, PhysRevE.88.042824, PhysRevE.100.052315, doi:10.1162/neco.2009.12-07-671, Goldman2009}.

In present work, the focus is the study of the dynamics of recurrent neural networks trained to perform Boolean-type operations with temporal stimuli at their input that {simulate sensory information represented in the brain}. In particular, the network's recurrent weights have been trained starting initially from matrices with weights given by a normal distribution with zero mean and variance $\frac{1}{N}$ by using backpropagation through time with the Adam method \cite{DBLP:journals/corr/KingmaB14}. { This will be explained in more detail in Section \ref{MODELO}.}

{In our previous work, whose preliminary version can be found in \cite{jarne2019detailed}, we have illustrated a set of properties of these networks. We studied how the performance degrades either as network size is decreased, interval duration is increased, or connectivity is damaged. We started a study on the eigenvalue's distribution, but without classifying the behavior. The main focus of such work was the connectivity properties related to the scale and network damage.} In the present work, the different aspects of dynamics have been studied in-depth. {An} interpretation is provided for the results of the numerical simulations corresponding to networks trained for the AND, OR, XOR temporal tasks, and a Flip Flop. Here the focus {is} the study of the eigenvalues spectrum of the recurrent weight matrices of trained networks and the classification of the obtained attractor states of the required output. {The aim is to show the link between the spectra in such tasks and dynamics.}

The motivation for the selection of these tasks is double. On the one hand, to simulate flow control processes that can occur in the cortex when receiving stimuli from subcortical areas. { In \cite{10.3389/fncom.2011.00001} the notion of gating was discussed as a mechanism capable of controlling the flow of information from one set of neurons to another}. { In the present work, the gating mechanisms are modeled using networks with a relatively small set of units}. On the other hand, these tasks are the basic and the lowest level for {computation} in any digital system. It has been previously proposed that some sets of neurons in the brain could roughly function as gates \cite{10.3389/fncom.2011.00001}. 

In the case of the Flip Flop, it is the simplest sequential system that one can build \cite{floyd2003digital}. {It is also interesting the dynamics of trained networks for the Flip Flop task, which is generally related to the concept of working memory}. It has been previously studied in \cite{DBLP:journals/neco/SussilloB13, SUSSILLO2014156}, but in this case, with a more complex task referring to a 3-bit register called in the paper a 3-bit Flip Flop.

So far, there are few detailed studies on the eigenvalues of the matrix of recurrent weights performed in trained networks. For example, the work of Rivikind and Barak \cite{PhysRevLett.118.258101} stands out. Although, the framework of this work is Reservoir Computing. Present work shares some of the observations made by the authors on the results. Other {previous} works considered matrices with partially random and partially structured connectivity, such as the {works} described in \cite{PhysRevE.88.042824, PhysRevE.91.012820, 10.1371/journal.pcbi.1006309}. There were also considered the results of these works in the present analysis.

Most of the existing literature on eigenvalues and dynamics is regarding the study of networks with random connections \cite{doi:10.1137/1129095, PhysRevLett.97.188104, PhysRevE.97.062314}. Besides, older works on dynamics, for example, considered other constraints such as symmetric matrices \cite{VIBERT1994589}.

For these reasons, the present analysis represents a significant contribution through the study of eigenvalues when considering non-normal matrices and trained networks. {The variability on the dynamics that can be observed on trained networks is surprising.} {The study presented here sheds light on the multiple solutions obtained for the same tasks and shows the links between the spectra of trained networks and the dynamics.} {Also, different properties of the model and the tasks were studied.}

The rest of the paper is organized as follows. The model and methods are presented in Section \ref{MODELO}. Section \ref{long} shows how to obtain the linearization of the system. In Section \ref{RESULTADOS NUMERICOS}, results are shown and how to classify the realizations obtained after training (network's simulations). In Section \ref{DISCUSION}, the different aspects that arise from the realizations and the dynamics are discussed. Finally, in Section \ref{CONCLUSIONS}, some remarks and further directions are presented.

\section{Model and methods} \label{MODELO}

The dynamics of the interconnected units of the RNN model is described by Equation \eqref{eq-01} in terms of the activity $h_i(t)$, where units have index $i$, with $i=1,2...,N$  \cite{Hopfield3088}.

\begin{center}
\begin{equation}
\frac{dh_i(t)}{dt}=-\frac{h_i(t)}{\tau}+\sigma \left( \sum_{j}w^{Rec}_{ij}h_j(t)+\sum_{j}w^{in}_{ij} x_j \right)
\label{eq-01}
\end{equation} 
\end{center}

$\tau$ represents the time constant of the system. $\sigma()$ is a non-linear activation function. $x_j$ are the components of the vector $\mathbf{X}$ of the input signal. The matrix elements $w^{Rec}_{ij}$ are the synaptic connection strengths of the matrix $\mathbf{W^{Rec}}$ and $w^{in}_{ij}$ are the matrix elements of $\mathbf{W^{in}}$ from the input units. The network is fully connected, and matrices have recurrent weights given from a normal distribution with zero mean and variance $\frac{1}{N}$, as already mentioned in Section \ref{INTRODUCCION}.

The readout in terms of the matrix elements $w^{out}_{ij}$ from $\mathbf{W^{out}}$ is:

\begin{center}
\begin{equation}
\mathbf{Z(t)}= \sum_{j}w^{out}_{ij}h_j(t)
\label{eq-02}
\end{equation}
\end{center}

For this study it was considered $\sigma () = tanh () $ and $ \tau = 1 mS$. The nonlinear function was chosen following models used in experimental works such as \cite{RUSSO2020745, WILLIAMS20181099, Remington2018}, but it could be changed without loss of generality, as well as the time scale. {The model is discretized through the Euler method for implementation} {folowing \cite{10.1371/journal.pcbi.1007655, 10.1371/journal.pone.0220547,Bi10530}}{, with a step time of 1 $mS$ obtaining Equation \ref{eq-03}, which is expressed in matrix form.} 

{
\begin{center}
\begin{equation}
\mathbf{H}(t+1)=\sigma(\mathbf{W^{Rec}}\mathbf{H}(t)+\mathbf{W^{in}}\mathbf{X}(t))
\label{eq-03}
\end{equation}
\end{center}
}

It was implemented in Python using Keras and Tensorflow \cite{chollet2015keras, tensorflow2015-whitepaper}, which allows making use of all current algorithms and optimizations developed and maintained by a large research community. {This procedure has previously been used in \cite {jarne2019detailed}.}

Networks were trained using backpropagation through time with the adaptive minimization method called Adam.
{ Initial spectral radius $g$, which is defined as the maximum absolute value of eigenvalue distribution of the matrix $\mathbf{W^{Rec}}$, is equal to 1. If the initial condition is far away from this value (with the proposed variance of $g\frac{1}{N}$), the training is not successful with few epochs. }

Although the training method is not intended to be biologically plausible, in a recent publication, arguments were presented regarding that, under certain approaches, this phenomenon could be plausible \cite{Lillicrap2020}. {The authors argued that the brain has the capacity to implement the core principles underlying backpropagation. The idea is that the brain could compute effective synaptic updates by using feedback connections to induce neuron activities whose locally computed differences encode backpropagation-like error signals}.

\begin{figure*}[htb!]
\begin{center}
\hspace{0cm}\includegraphics[width=16cm]{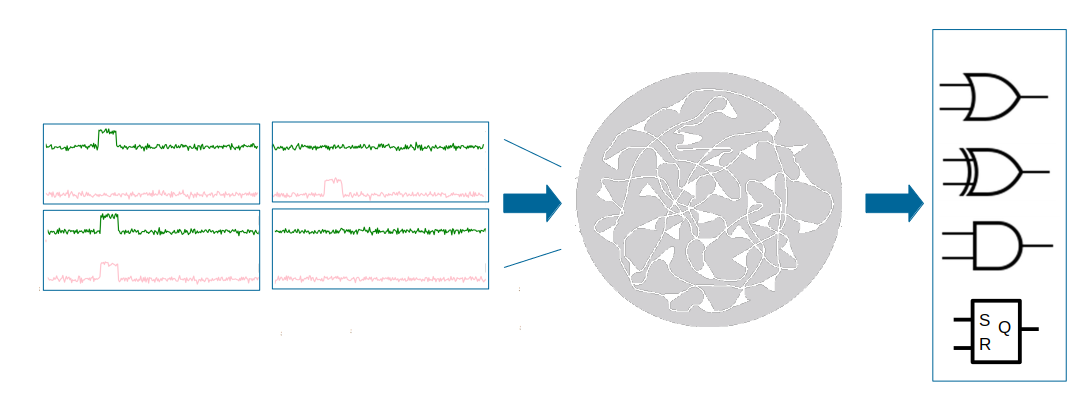}
\caption{Model. In the training stage, the time series are introduced into the network in the 4 possible combinations constituting a set with 15000 samples with noise. The training algorithm adjusts the weights according to the target function, to obtain the trained matrices $ \mathbf{W^{in}}$, $ \mathbf{W^{Rec}}$ and {$ \mathbf{W^{out}}$ for each task.} \label{fig_01}}
\end{center}
\end{figure*}

The stimuli presented at the input of the networks, corresponding to the training sets, are time series containing rectangular pulses with random noise corresponding to 10 \% of the pulse amplitude. The possible combinations presented are: 2 simultaneous pulses at each input, {a single} {stimulus} in one or the other, or no pulse, constituting the four possible binary combinations. {These are shown} on the right side of Figure \ref{fig_01}, where also a representation of the model is presented. The target output completes the set, and it depend on which of the functions {is selected} to teach the network (AND, OR, XOR, or Flip-Flop).

Networks of two different sizes were considered for the study: 50 and 100 units, the latter as a control case. The tasks can be learned in reasonable computational time, and with good accuracy, with 50 units.  {Two types of initial conditions were considered for the recurrent matrices: Random Normal distribution and Random Orthogonal \cite{jarne2019detailed}}. The second case is an additional constraint. {It means that the matrix} is initialized with an orthogonal matrix obtained from the decomposition of a matrix of random numbers drawn from a normal distribution.

Although successfully trained networks can also be obtained using the identity matrix for initialization, this initial condition is far from the random connectivity paradigm previously used.

Networks were trained to carry out all the mentioned tasks (AND, OR, XOR, and Flip Flop). {More than 20 networks were trained for each initial condition and task.} {Simulations and code to perform the analysis of present work are available in Appendix \ref{sup-c}.} 

{To summarize, Table \ref{tabla1} contains the parameters considered, such as the size of the network, data set, noise and regularization terms that are appropriate for the considered task.}

\begin{center}
\begin{table}[htb!]
{
\begin{tabular}{|l|l|}
{Parameter/criteria}                   & Value          \\
\hline
Units                                  & 50               \\
Input Weights                          & $2\times50$      \\
Recurrent Weights                      & $50\times50$    \\
Output Weights                         & $50$      \\
Training algorithm                     & BPTT ADAM         \\
Epochs                                 & 20                 \\      
Initialization                         & Ran Orthogonal$-$ Ran Normal \\
Regularization                         & None  \\
Input Noise                            & 10\%
\end{tabular}
}
\caption{{Model's parameters and criteria for the network's implementation and training.}}
\label{tabla1}
\end{table}
\end{center}

\begin{figure*}[htb!]
\begin{center}
\includegraphics[width=15cm]{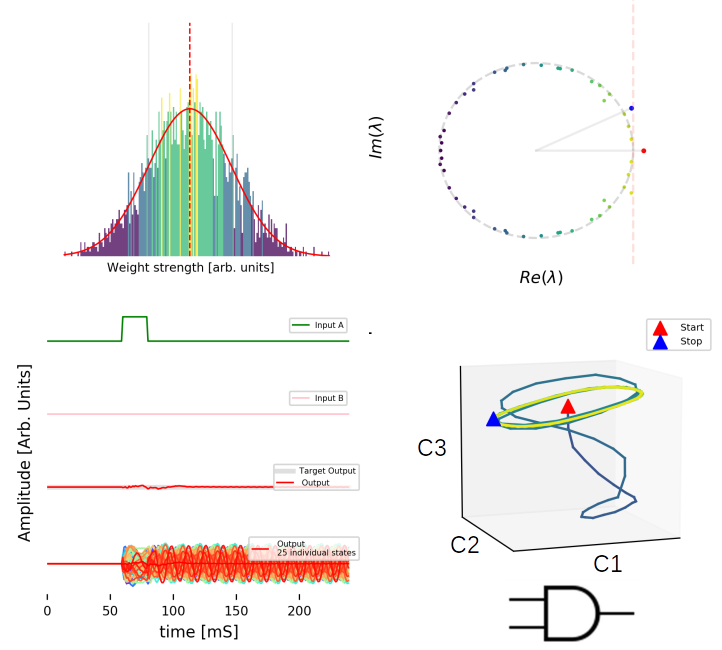}
\caption{ {Methods. Panel a) shows the weight distribution of $ \mathbf{W ^{ Rec}}$. Panel b)  shows the eigenvalue distribution in the complex plane, corresponding to the decomposition of the $\mathbf{ W ^ {Rec}} $ matrix. Panel c) is a possible combination of stimuli (High-Low) presented to the network with its output and the temporal response of some units. Panel d) presents the decomposition into singular values in the 3 main components (C1, C2, C3) or axes, performed for the 50 unites states $h_k(t)$ and the temporal evolution of the main components shown for the considered period.}\label{fig_02}}
\end{center}
\end{figure*}

For each realization, the distribution of the recurrent weights pre and post-training was plotted. The distribution moments were estimated in each case. Then, the decomposition of $ \mathbf {W ^{rec}} $ in their eigenvectors and eigenvalues was obtained. An example of one network is presented in the upper part of Figure \ref{fig_02}. {In panel a), the distribution of the weight matrix is shown. In panel b), the distribution of the eigenvalues in the complex plane is presented}. The behaviour is described in detail in Section \ref{dist_autovalores}.

The realizations obtained were studied and classified one by one. To do this, a noise-free testing set, corresponding to the four possible binary options, was used to study the response of each network. The behaviour of some k units was plotted as a function of time for each of the possible stimuli combinations. The {panel c)} of Figure \ref{fig_02} shows the response of the set of units $ h_k (t) $, corresponding to a network trained for the AND task with a stimulus at one of its inputs. In this case, input A is elicitated. {After the stimulus of one input only,  the network's output must remain in a "Low" state as expected, since in the task AND the output goes to a "High" state only when both inputs receive a stimulus.}

{A decomposition into singular values was performed with the entire set of the output's units $ h_i (t) $}. {It was done for each input combination, using the single value decomposition method from the scientific library Scickit Learn \cite{scikit-learn}. These are} {Python open source libraries based on Numpy that allows us to perform dimensionality reduction, feature extraction, and normalization, among other methods for predictive data analysis.} 

{The behaviour of the system was plotted into the 3 axes of greatest variance. This is shown in {panel d)} of Figure \ref{fig_02}, for one example.}

\section{Long term dynamics} \label{long}

To interpret the results obtained in the realizations, {which were classified and presented} in the following sections, let us begin by making some approximations regarding the system. These will allow us to understand the behaviour of the $ h_i (t) $. {In this way, we can perform one analysis of the connectivity matrix and another on the activity.}

If the units operate away from the saturation regime, we could perform a linearization of the system to make an approximate description of the long-term dynamics. That will allow us to associate our observations with some well-known results.

From Equation \eqref{eq-01} we can consider the linear model given by Equation \eqref{eq-lin}, using the first order Taylor expression for $tanh()$.

{
\begin{equation}
\frac{dh_i(t)}{dt}=-h_i(t)+\sum^N_{j=1}w^{Rec}_{ij}h_j(t)+\mathbf{I(t)}.w_{0,i}
\label{eq-lin}
\end{equation}
}

In the absence of external input, the system has a single fixed point that corresponds to $ h_i = 0 $ for all units $i$. We can write the external input as a time variable component $ \mathbf {I (t)} $ and a term $ w_ {0, i} $ that corresponds to the activation of each unit. Let us then consider a vector $ h_ {0} $ N-dimensional, and let's approximate the input pulse $ \mathbf { I (t)} $ by the delta function. {This} means that the duration of the pulse is short relative to the length of the time series considered, as is {in} our case.  In addition, the norm of $ w_ {0} $ is 1, which is equivalent to saying $ h (0) = w_0 $.

The solution, given by the Equation \eqref{eq-lin} {and} following \cite{9910223127702121,10.5555/1205781,10.1371/journal.pcbi.1007655}, is obtained by diagonalizing the system and making a base change of the vector $ \mathbf{h} $, such that:

\begin{equation}
\mathbf{ h }=\mathbf{V} \mathbf{\tilde{h}} 
\end{equation}

Where the columns of $\mathbf{V}$ are the eigenvectors of $\mathbf{W}$. Then, it is possible to write the connectivity matrix, $ \mathbf {W^{Rec}} $, in a diagonal base containing the eigenvectors $ v_i $ as columns, and the matrix $ \mathbf {\Lambda} $ that has the eigenvalues $ \lambda_i $ on the diagonal as shown in Equation \eqref{eq-lin-2}.

\begin{equation}
\mathbf{W^{Rec}} \rightarrow \mathbf{\Lambda}=\mathbf{V^{-1}W^{Rec}V}
\label{eq-lin-2}
\end{equation}

This is used to decouple the equations. Now we {can} write the decoupled equations of $ \tilde {h} _i $ for the vector in the new base as in Equation \eqref{eq-lin-3}:

{
\begin{equation}
\frac{d\tilde{h}_i(t)}{dt}=-\tilde{h}_i(t)+\lambda_i\tilde{h}_i+\delta(t).\tilde{w}_{0,i}
\label{eq-lin-3}
\end{equation}
}
In this way we obtained the solution for $ h $ in terms of the $ \tilde{h_i}$

\begin{equation}
\mathbf{h}(t)=\sum^N_{i=1}\tilde{h}_i(t)\mathbf{v_i}
\label{eq-lin-4}
\end{equation}

with
\begin{equation}
\tilde{h}_i(t)=e^{t(\lambda_i-1)}
\label{eq-lin-5}
\end{equation}

Thus, the long-term dynamics is governed by the eigenmodes with the eigenvalue (or eigenvalues) with the largest real part. 

{Linearization performed in this section has some limitations to interpret the results of trained networks. In particular, when eigenvalues satisfy that $Re(\lambda)>1$. Considering the linearization, one would expect that the system diverges with exponential growth, but in the case of trained networks it won’t. The reason is that, by design, the network is trained for a task, then the activity can not diverge to archive the considered task. In the end, the non-linearity conspires in favour. Based on this situation, the eigenvalues outside the unit circle are finally associated with oscillatory or fixed-point behaviours. The hyperbolic tangent flattens the amplitude of the activity in such cases, and the units activity reaches a fixed value, which is given by the hyperbolic tangent of the product of the amplitude of eigenvectors and the exponential function. When eigenvalues are complex, there are oscillations, as we observed with the frequency given by Equation \ref{eq-lin-6} in Section \ref{linking}.}

This is true for all the realizations obtained in this work since the output's state always corresponds to one of the responses to the combinations of the stimuli, being the active or passivated, oscillatory, or fixed-point output.

\section{Results \label{RESULTADOS NUMERICOS}}

{The results of the numerical studies are shown in this section. The first study is regarding the statistics on the recurrent weight matrix and is presented in Section \ref{statistics}. The second is on the activity of trained networks (shown in Sections \ref{clas-macro} and \ref{classification}).} { The third is on the distribution of eigenvalues (Section \ref{dist_autovalores}), where it is shown the link between eigenvalue distribution and dynamics considering the analysis presented in Section \ref{long}. Finally, Section \ref{linking} shows more details on the dynamics infered from the recurrent weights matrix. }

\subsection{{The distribution of the weight matrix of trained networks}} \label{statistics}

It is possible to compare the differences in the distribution of the weight matrix of trained networks by studying the pre and post-training moments. {The changes between the initial and final states of the parameters of the distributions during training were studied using linear regression. The variation with respect to the identity function (meaning no parameter’s change) was considered. The percentage variations are reported in Appendix A.}

{It was observed that the variation of the post-training mean value is less than \% 6 for all the tasks, with a tendency to decrease with respect to the initial condition.} Regarding the standard deviation, the variations are less than 0.5\%. In the case of Skewness and Kurtosis, they increase slightly by a maximum of 15 \% in the worst case, and in the case that least varies, the variation is less than 0.5 \%.  { For tables and full details, see Appendix \ref{A}.}

{The distributions of the recurrent weights matrix of trained networks do not differ significantly with respect to the pre-training ones. Therefore, studying the differences purely in the parameters of the distributions due to the training does not yet allow us to understand aspects related to dynamics, which also motivated the study of the eigenvalue distribution presented in Section \ref{dist_autovalores}.}

\subsection{ {Classification of the tasks}}\label{clas-macro}

{AND, XOR and OR tasks considered in this study compute temporal stimuli. They are different from the binary operations in feed-forward networks. These tasks presented here are a class of decision-making tasks and are considered for the first time to describe their dynamical behaviour.}

{From inspecting the different realizations obtained, some general observations of these systems emerge when tasks are compared. It is possible to group the tasks with respect to the number of minimum modes or states that the system needs to have to accomplish the tasks.  AND and XOR as similar tasks, OR as a simpler one, and Flip Flop as a slightly more sophisticated task related to AND and XOR.}
 
{First, let us consider the case of the AND and XOR tasks. There are three general states of the system for both. Each state is shown in Figure \ref{fig_03}. The resting state is shown in panel c). A second state, where the stimuli at the inputs produce a high-level state at the output, is presented in panels a) and b). The other, where the stimuli elicited activities $h_i(t)$ that produces the output's passivated state despite the stimuli at the inputs, is shown in panel d). The difference between the AND and XOR task is which combination of stimuli at the inputs elicitate each output state.}

{The OR task is simpler, in the sense that for any combination of stimuli presented at the input, the state of the output must be high-level. In the case of not having a stimulus at any input, the output must be zero. For this task, there is no combination of stimuli that leads the output to be passivated, as in the case of the AND and XOR functions. There are only two possible general states for the system: The resting state and the state that activates the output.}

In the case of the 1-bit {Flip-Flop}, one stimulus at the input called "S" brings the system to the high-level state, while another in {input} "R" takes it to the passivated state. {Two consecutive stimuli at input "S" or input "R" does not generate changes in the system. This task is more complex since the changes depend on one specific input. It is necessary to consider also that the system has to remain in the same state when applying two consecutive stimuli, meaning that the system must ignore a second activation of the same input of each one.}

{It is possible to summarize these ideas as follows: AND and XOR need to have at least three modes. Two general modes, in terms that each state of $h_i(t)$ elicited by the input} associated with the possible states of the system, and other for the resting state. The same is true for the Flip Flop, which also needs to remain unchanged when consecutive stimuli {are applied}. OR task needs to have two modes at least. One is associated with the high-level state, and the other is the resting state.

{The classification made corresponds to the observation that certain binary decisions require a passivated state while others do not. This is related to the eigenvalue distribution shown in Section \ref{dist_autovalores}.}

 {When training the networks for AND, OR, XOR, and Flip Flop tasks, similar configurations arise for the distributions of the eigenvalues. They will also be described in more detail later in Section \ref{dist_autovalores} and is related to the minimum number of eigenvalues outside the unit circle.}

 {It is observed that for the same task} {it} {is not unique how each network manages to maintain the state of the output for which it was trained, as we have previously observed in \cite{jarne2019detailed}. There are different ways to combine the network weights to have different internal states that result in the same required output rule. These lead to different behaviour in the dynamics.} {This is shown in Section \ref{classification} and discussed in more detail in Section \ref{DISCUSION}.}

\subsection{Classification of the realizations} \label{classification}

{Section \ref{clas-macro} allows classifying the obtained realizations into different groups. In this section, we will observe the variations in the activity that occurs within the same class of tasks.} {The behaviour of the activity $ h_i (t) $ is used to understand what happens when each network's input is elicitated with the four possible different combinations at the inputs.}

\begin{figure*}[thb!]
\hspace{0cm}\includegraphics[width=17cm]{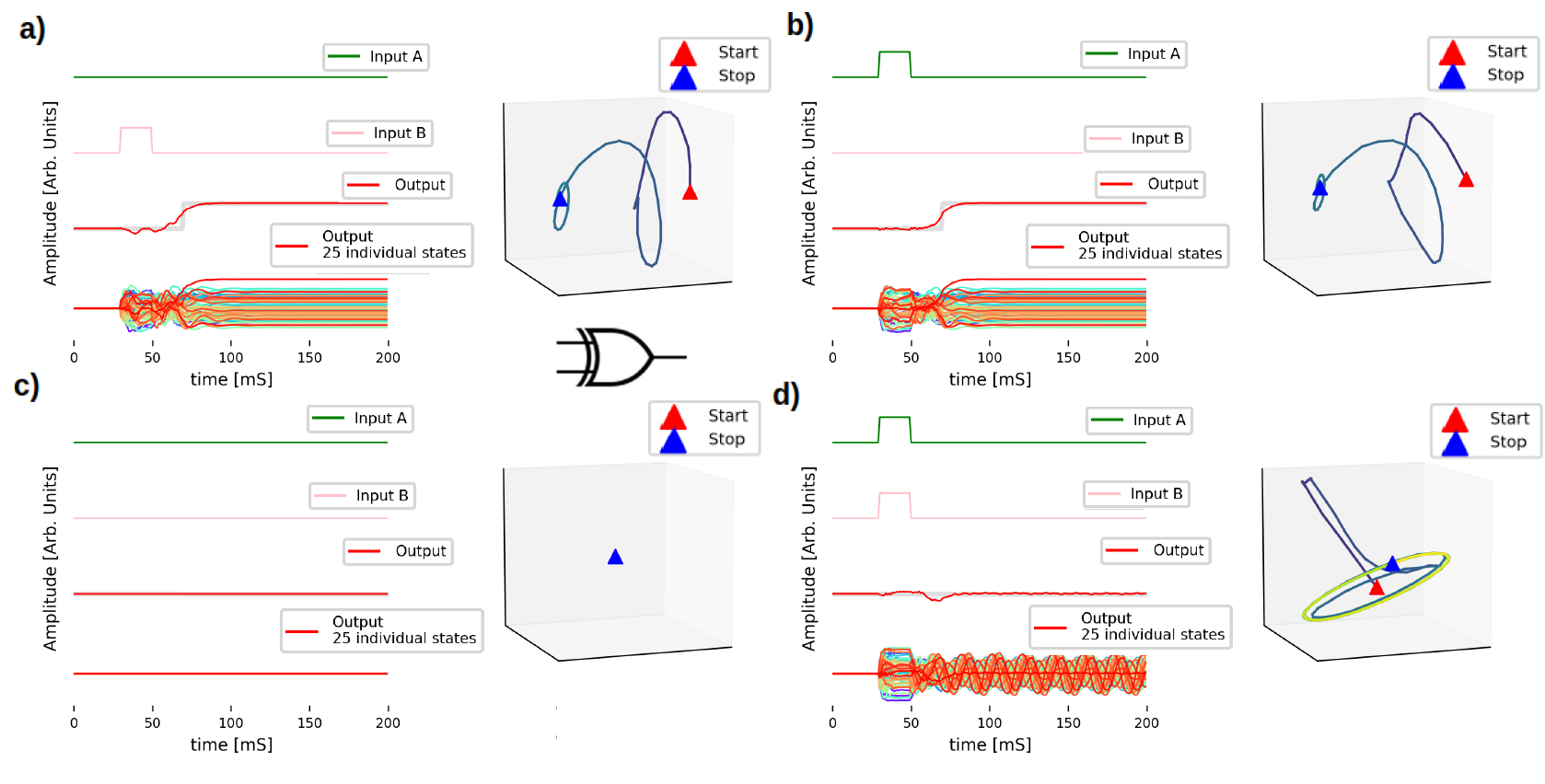}
\caption{{Example of the four different stimuli for the XOR task. Upper panels a) and b) show high-level output states (fixed-point states) for either input stimulus. Panel c) shows the resting state. Panel d) presents a passivated oscillatory state of the output in response to the presence of two simultaneous stimuli. These results correspond to realization with label $XOR\ \#id10\_Ortho$ in the dataset.\label{fig_03}}}
\end{figure*}

{The trained networks have been labelled to follow the examples shown in this section and the next one.} { The labels include the name of the task for which the network was trained, a number to identify the realization, and the initial condition of the weight distribution. These labels allow also identifying the data in the Supplementary Information to view additional examples and to have the corresponding raw data available. For example, "$XOR\ \#id10\_Ortho$", shown in Figure \ref{fig_03}, is an example of a network trained for the XOR task, with the orthogonal initial condition and the id number 10.}

{Let's start describing the activity in the case of the AND and XOR tasks.} First, the passivated output mode is described. The following situations may occur:

\begin{enumerate}

\item When the stimulus arrives, the $ h_i (t)$ activities start a transitory regime that ends in a sustained oscillation, each with a different phase and amplitude. The superposition is given by $ \mathbf {W ^ {out }} $ and allows to passivate the output. An example of this {behaviour} is shown in { panel c)} of Figure \ref{fig_02}.

\item When the stimulus arrives, $ h_i (t) $ start a transitory regime that leads to a fixed level other than zero for each unit, and whose superposition, given by $ \mathbf {W ^ {out}} $, allows to passivate the output.

\item When the stimulus arrives, $ h_i (t) $, passes to a transitory regime that attenuates to zero, and the output is null as a result of the attenuation of the $ h_i (t) $ activity.

\end{enumerate}

{If we consider now the mode of the excited output-state, we have situations similar to 1) and 2), leading to a high-level output state, but is not possible to have 3). Also,} it is observed in the numerical simulations that the sustained oscillatory mode is more often associated with the passivated state of the output, as shown in Figure \ref{fig_02} (\textbf{See also Supplementary Information}).

Let's illustrate this situation with the realization with label $XOR\ \#id10\_Ortho$, represented in Figure \ref{fig_03}. {In this case,} the excited output-state appears as a fixed point final state, while the passivated output appears as an oscillatory state.

The possible combinations listed above correspond to the observation for the different realizations. { Thus} it is possible to have either an excited state with oscillatory behaviour for the $h_i(t)$ or an excited state with a fixed point (\textbf{See Supplementary Information}). The same is true for the passivated state of the output.

Now let's consider the OR task. In this case, there is {only} one active mode corresponding to any combination of stimuli. The situations that can occur are:

\begin{enumerate}

\item With any stimulus of the inputs, $ h_i (t) $ passes from a transient to a fixed point.

\item With any stimulus of the inputs, $ h_i (t) $ goes from a transient to a sustained oscillation regime.

\end{enumerate}

{The case of zero output for the OR tasks corresponds only to zero stimuli at the inputs. Additional examples for this task are shown in the \textbf{Figure 2 of Supplementary information} illustrating all stimuli situations.}

\subsubsection{A second stimulus}\label{one second}

{The networks of each task have not been trained to respond in a particular way to a second stimulus (or stimuli) temporarily delayed from the first.}

{A study was carried out to test if it is possible to generalize the behaviour of the trained networks when they receive a second stimulus at the input. It was considered two cases: identical stimuli to the first one, and opposite stimuli. The case of two identical stimuli was studied with the aim to link the response of the trained networks with the Flip Flop task. In the second case, the opposite stimulus was used to test if it can act as a restitutive response of the system to return to the initial state.}

{We have observed that the behaviour cannot be fully generalized, however, useful observations from this study have emerged and are presented in this section.}

If, after a certain time, the network receives a second stimulus equal to the previous one (in one or both inputs), it is possible to classify the response of the system according to which was the previous input state and what is the task for which it was trained. 

\begin{figure*}[htb!]
\hspace{0cm}\includegraphics[width=17cm]{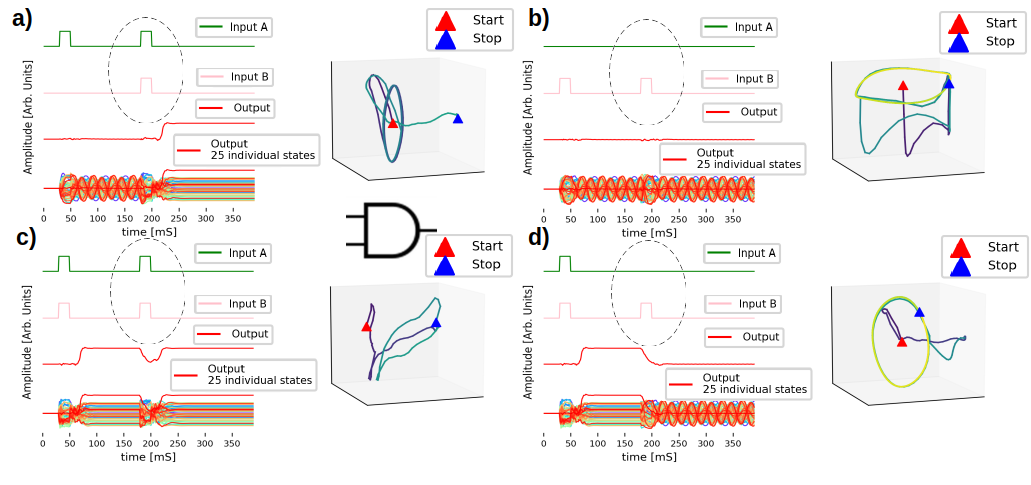}
\caption{{Trained network for the AND task corresponding to the realization with label $AND\ \#id15\_Ortho$ considered in Figure \ref{fig_02}.} {The network is elicitated with a second stimulus in one or two inputs, identical to the first one. Each of the panels shows the different relevant situations described in Section \ref{one second}.\label{fig_05}}}
\end{figure*}

{For example, let's consider the situation where the network is trained for the AND task and presents the passivated output state, such as the example considered {in panel c)} of Figure \ref{fig_02}. In the case of receiving a second stimulus at both inputs, the network migrates to a new state, so the output goes to a high-level state (as seen in panel \textbf{a)} of Figure \ref{fig_05}), where the additional stimuli are indicated with a dashed circle. If it receives a single second stimulus, the system is disturbed, but it returns to the passivated condition (generally an oscillatory state) so that the output is set at zero level, as seen in panel \textbf{b)} of Figure \ref {fig_05}.}

Now let's consider the case where the output is in a high-level state, and the system receives two simultaneous stimuli again. In this case, the system is disturbed, but it remains at the high-level state, as shown in panel \textbf{c)} of Figure \ref{fig_05}. If the network receives a new stimulus  (in one of the inputs only), the state to which it migrates depends on each particular realization, and it is not possible to classify the response in a general way. For the realization shown in panel \textbf{d)} of Figure \ref{fig_05}, the system goes to the passivated state.

{In all cases, it can be observed how the trajectory in the low-dimensional space defined by the 3 axes with the highest variance becomes more complex when the second input stimulus is induced.}

{To train a network for the Flip Flop task, it is necessary a configuration in which its response is similar to Figure \ref{fig_05} b) when two consecutive stimuli at the same input are presented. Indeed, when reviewing the realizations obtained for said task, they behave like those AND implementations that present a response like the one in Figure 4 b).}

{We can analyze if there is any restorative response of the system when considering a second stimulus at the input with the opposite sign to the first one. Again, the network has not been trained to respond in a particular way to this situation.} 

\begin{figure*}[htb!]
\hspace{0cm}\includegraphics[width=17cm]{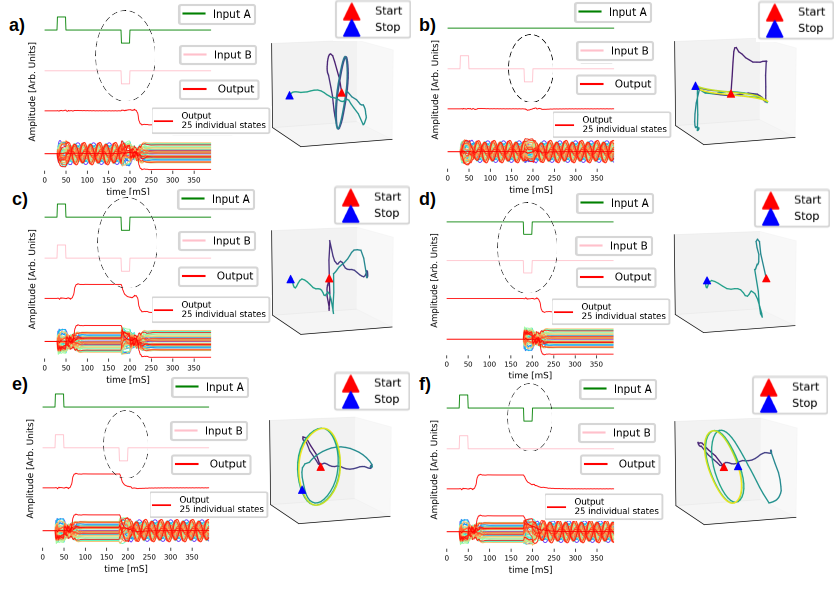}
\begin{center}
\includegraphics[width=8cm]{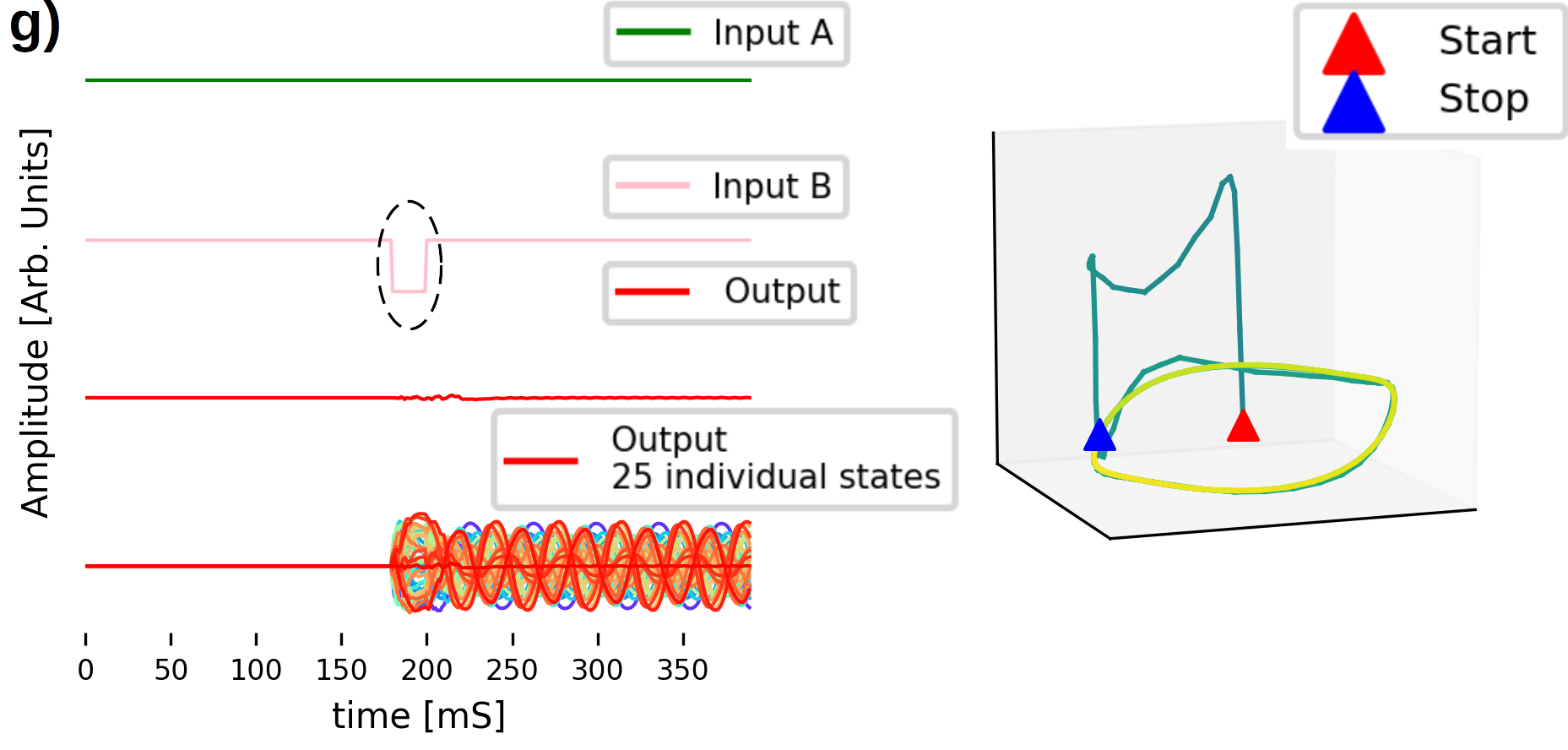}
\end{center}
\caption{{Trained network for the AND task considered in Figure \ref{fig_02} with a second negative stimulus in one or two inputs (corresponding to the realization with label $AND \#id15\_Ortho$). Each of the panels shows the different relevant situations described in Section \ref{one second}, and the behavior of the system in low dimensional space according to each case.\label{fig_06}}}
\end{figure*}

If the network receives the second stimulus with the opposite level of the first one (in one or two of its inputs), it is possible again to classify the response of the system according to the previous state. {This is illustrated in Figure \ref{fig_06}, where it is considered again the AND task}. One possible state is to have the output at a low-level, corresponding to the passivated state produced by a single previous stimulus (panels \textbf{a)} and \textbf{b)} of Figure \ref{fig_06}). As shown in panel \textbf{a)}, if the network receives two stimuli, the output migrates to a negative level. If it receives a single negative stimulus, the system is disturbed, but it remains in the passivated state, shown in panel \textbf{b)} of Figure \ref{fig_06}.

Now, lets consider {the output of} the system  being in a high-level state when receives one negative stimulus (panels  \textbf{e)} and \textbf{f)}) or two negative stimuli at the input (panel \textbf{c)}). In both cases, the state of the output depends on the realization, and it is not possible to classify the response in a general way.

If the network output is at a low level and receives two negative stimuli at the inputs, it migrates to a negative state. This case is shown in panel \textbf{d)}. If the network receives a single negative stimulus, it migrates to the passivated state, shown in the lower central panel of Figure \ref{fig_06} {(panel \textbf{g)}).}

{Finally, we achieved passivated states only in the case of stimulating a single input with a negative pulse, while when the stimulus is in both inputs, the output migrates to a negative state.}

{In this way, producing a stimulus of the opposite sign at the inputs, as shown in the cases of Figure 5, does not allow the system to return to the state before the first stimulus (restorative response), but rather it takes it to a new different state or disturbs it and continues in the same state as panel \textbf{b)} of Figure 5.}

\subsubsection{The Flip Flop}\label{FF}

 {The case of the Flip Flop is different and interesting because the dynamics of trained networks for the Flip Flop task is related to the concept of working memory.} The Flip-Flop {activity} is also more difficult to analyze, {because the output state is designed to be transitory}. However, when observing the response of the AND networks to a second positive stimulus, it is possible to detect the different situations that could arise in favor of having a Flip Flop { (See Figure \ref{fig_05}). Depending on the input stimuli combination that follows the first, the output state could migrate to a low level or remain in high state. The low-level state can correspond to either a fixed point or a sustained oscillation of the activity $h_i(t)$. This state will continue until the next stimulus on the inputs changes its value again}. When system remains in the seame state it is possible that the stimulus on the same input disturbs it a little with noise but {still} allows the system to maintain a sustained state. 

\begin{figure*}[htb!]
\hspace{1cm}\includegraphics[width=14cm]{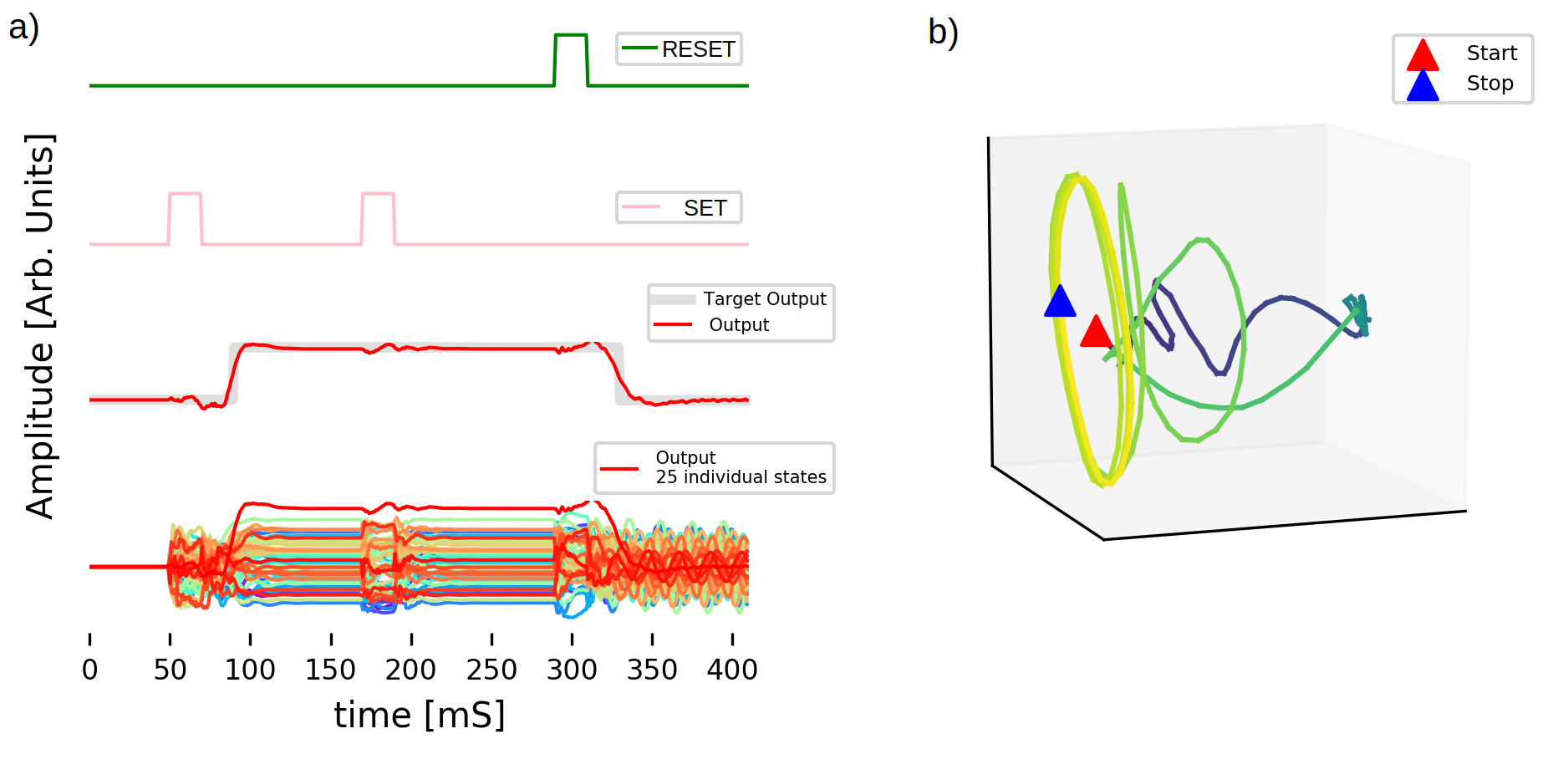}
\caption{Example of one simulation performed with a trained network for the Flip Flop task. This case corresponds to the network with label $FF\ id\# 05\_Ortho$. {Panel a):} the state of the SET and RESET inputs are shown as a function of time. The outputs and the temporal evolution of the activity of some units are also shown. {On the right side, panel b), the behaviour of the system was plotted into the 3 axes of greatest variance.}\label{fig_07}}
\end{figure*}

This situation is shown in Figure \ref{fig_07}, where the output result is presented for one of the realizations, corresponding to the network with label $FF\ id\# 05\_Ortho $. Here it is shown two consecutive stimuli at the "S" Input, and then another at the "R" input.

In the Flip Flop task it is necessary that, when stimulating the "R" input, the system migrates to a fixed-point or an oscillatory state, corresponding to the passivated output state. By stimulating the "S" input, the system must migrate {similarly} to an active state. The system must also have a mechanism that allows ignoring consecutive stimuli.
{The realizations obtained for this task show a great diversity of behaviours \textbf{(See Supplementary Information)}.}

\subsection{The eigenvalue distributions of the realizations}\label{dist_autovalores}

There are regular patterns in the distribution of the eigenvalues of the recurrent matrix {for all considered tasks}. This situation happens for the matrices of the trained networks that have been initialized before training with the random normal condition as well as for that trained starting from the orthogonal random normal condition.

These patterns can be characterized. {They are very similar to the initial condition (pre-training), but with a set of eigenvalues outside the unit circle.}

\begin{figure*}[htb!]
\hspace{1cm}\includegraphics[width=7.5cm]{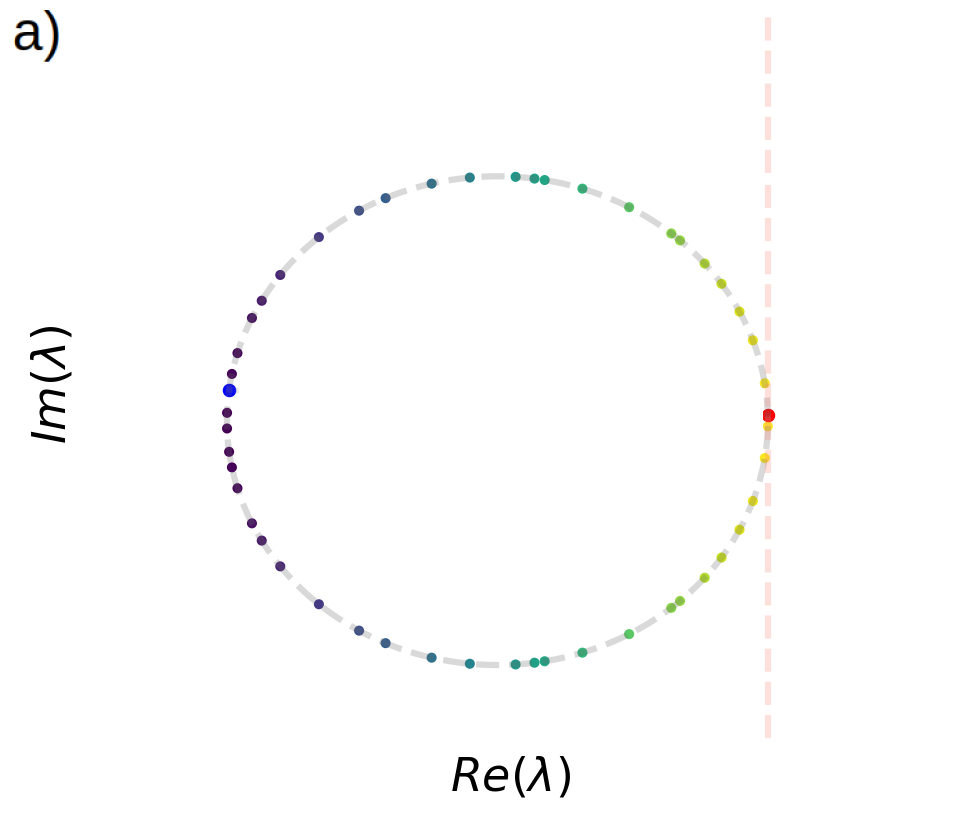}
\includegraphics[width=7.5cm]{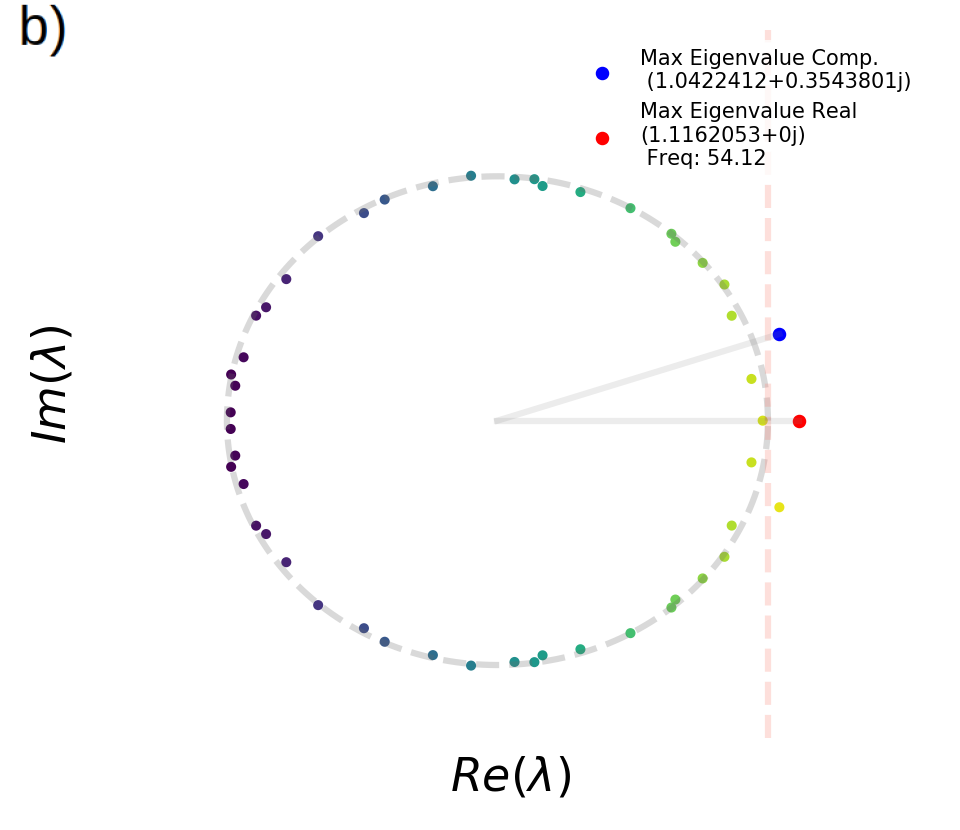}
\caption{Comparison between the distribution of eigenvalues corresponding to the pre-training and post-training condition for the network previously considered in the example shown in Figure \ref{fig_03}. On panel a), it is shown the orthogonal condition reflected in the distribution of the eigenvalues. On panel b), it is shown the result that after training. A few eigenvalues migrate out of the unit circle. \label{fig_08}}
\end{figure*}

Let's consider, for example, the initial condition of the case previously presented in Figure \ref{fig_03} of the XOR function, and let's compare it with the trained network. This is shown in Figure \ref{fig_08}.

The figure shows that, except for a small group of eigenvalues that migrated out of the unit circle, the rest remain inside the unit circle. This situation is repeated in all the simulations obtained \textbf{[See Supplementary Information]}. From this observation, it is proposed that the eigenvalues outside the unit circle are directly related to the modes of $h_i (t)$ that configure the possible states of the output. We have suggested it previously in \cite{jarne2019detailed}, which is also compatible with the observations made in \cite{doi:10.1162/neco.2009.12-07-671}.

The location of the eigenvalues outside the unit circle seems to be related to the behavior (or mode) observed for the different stimuli discussed in the previous section. Indeed, for all the realizations obtained corresponding to the different tasks, it was possible to link the position of the eigenvalues with the approximate behavior of the unit's activity $ h_i (t) $, described in the previous section.

\begin{figure*}[htb!]
\hspace{0cm}\includegraphics[width=18cm]{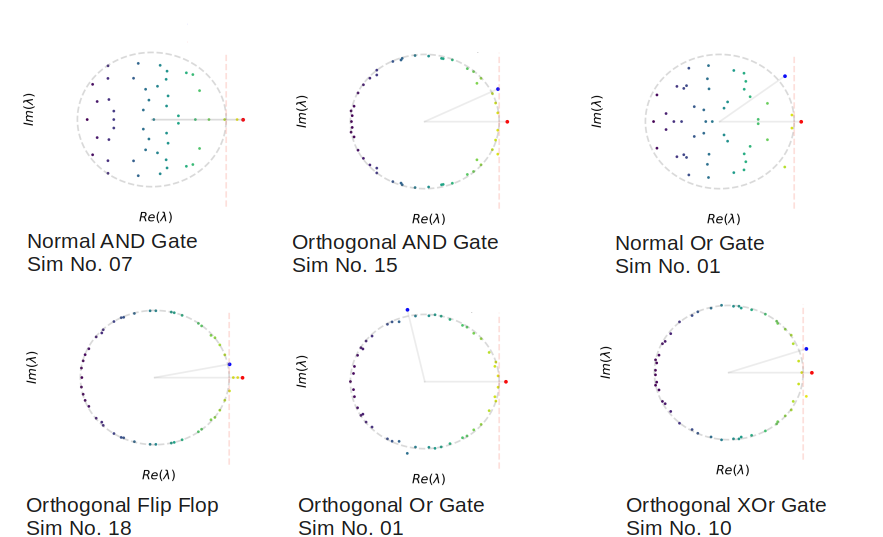}
\caption{ {Distributions of eigenvalues in the complex plane for the realizations used to exemplify the different modes obtained as a result of training and initial conditions. It is observed that the dominant values outside the unit circle can be real or complex.} \label{fig_09}}
\end{figure*}

In Section \ref{long}, it is argued why the analysis of the recurrent weights matrix allows a good approximated description of the different modes obtained for each realization and stimulus type. Now, let's classify the different distributions of eigenvalues of the realizations, and relate them to the results presented in Section \ref{classification}.

{The eigenvalue distributions for different tasks and initial conditions of the most common situations obtained in the simulations are presented in Figure \ref{fig_09}.}

Let's consider the AND and XOR tasks. It is observed for these tasks that the $\mathbf{W^{Rec}}$ matrices present at least 3 eigenvalues outside the unit circle in more than 70\% of the realizations obtained. One usually is a real eigenvalue, and the others constitute a complex conjugate pair. Different cases can occur in this frequent situation. Those are described below [\textbf{See also Supplementary Information}].

The fixed level of activity $h_i(t)$ is usually associated with the excited level of the output, while the complex conjugate pair usually is associated with the passivated level. Exceptionally, it is possible to observe a few cases where this is the other way around (less than 20\% of cases, \textbf{See Supplementary Information}). It is also observed that the frequency of oscillation of $ h_i(t)$ always correlates with the angle in the complex plane defined by Equation \eqref{ang}.

\begin{center}
\begin{equation}
 \theta= arctan \left(\frac{Im(\lambda_L)}{Re(\lambda_L)}\right)
\label{ang}
\end{equation}
\end{center}

$\theta$ is measured with respect to the positive semi-axis, $\lambda_L $ is the complex dominant eigenvalue outside the unit circle (imaginary part is not zero). Small angles correspond to slower frequency oscillations of the activity $ h_i (t)$, while larger angles correspond to faster oscillations, as is shown also in \textbf{[Figure 1 of Supplementary Information]}. 

In approximately 10\% of the realizations, the eigenvalues outside the unit circle are pure reals (a rare situation where there are usually 2 or 3 eigenvalues outside the unit circle). {The} states of the $ h_i (t) $ correspond to non-zero sustained fixed levels. This happens for both states, {the passivated output and the excited output.}

When the eigenvalues outside the unit circle are 2 pure reals, but one is on the side of the negative semi-axis, a fixed-level mode appears for the $ h_i (t) $ and another mode with very fast oscillations \textbf{[See Figure 2 of Supplementary Information]}.

Exceptionally (in less than 10\% of the realizations), some trained networks have more than two pairs of complex conjugates. In this case, the oscillatory behaviour is usually more complex, but it seems to be dominated by the eigenvalue more distant from the unit circle. {In cases of high-frequency oscillations, slow modulations can also be observed in the levels of $ h_i (t)$, on those networks that have negative real eigenvalues or eigenvalues with large imaginary parts.}

Let's consider now the results obtained for the OR task. In this case, as mentioned in Section \ref{classification}, it is enough to have one general mode for the activity of the units, since it is possible to either have the state of rest or the excited state of the output. There is no passivated state in this task.

Due to the stochastic nature of the training algorithm, as well as { the differences in} the initial parameters, multiple final configurations for the trained network are possible for the same task. However, there is a minimum requirement in the connectivity of the network to perform the task. { For the OR task, it is necessary to have at least one eigenvalue with a real part greater than 1 in the recurrent weights matrix for the network to be able to perform the task. This eigenvalue corresponds to the high-level excited output state.} It is possible to have a realization with more than one mode, but this will not be elicited by {any} combination of stimuli at the input.

In the case of matrices with the initial condition orthogonal, the configurations mostly have 3 eigenvalues outside the unit circle: the complex conjugate pair and the pure real eigenvalue. {For} random normal matrices, it is most common to have only one pure real eigenvalue.

This difference between both conditions appears because when the eigenvalues are located on the edge of the circle (orthogonal initial condition), it is less difficult for the training algorithm to move a complex conjugate pair outside the unit circle. All are placed at the same distance on the border. When a training instance happens, all eigenvalues change a little their position. Whereas, if the initial condition is random normal, it is more computationally expensive to push more {than} one eigenvalue, since they are more likely located further from the edge at different distances, and more instances of training are necessary. But if the matrix were initialized with some of the eigenvalues out the border, such as the case of the simulation with the normal condition for the OR task (Sim. 01) shown in Figure \ref{fig_09} upper right panel, it is possible to have a realization with more than one mode after training. For complete detail of simulations see \textbf{the statistical summary of Supplementary Information}.

Depending on the proximity to the edge, it is possible to have configurations with a single mode or two. In the case of having two, the stimuli generally elicitate the mode corresponding to the pure real eigenvalue. {The activity} $ h_i (t)$ {goes} from the transitory state to the fixed level, which is consistent with the previous observation for the AND and XOR tasks, where the oscillatory states usually {correspond} to the passivated output level, {which is} a state that does not occur for any combination of stimuli in the OR task.

{This result has been previously observed in the simulations of \cite{jarne2019detailed}, but not described.} It is known from the Machine Learning field that orthogonal initialization of the weight matrix in recurrent networks improves learning \cite{pmlr-v48-arjovsky16}. {It is used sometimes as a regularization technique when minimizing the gradient, among others.} 

Let's consider the Flip Flop task. For this task, the minimum situation for the system {to fulfil} the task is analogous to what happens in networks that learned AND or XOR tasks. For a given combination, the network must be able to {present} the passivated state of the output. 

The cases obtained in this work can be classified into similar categories as before. Nevertheless, this task has an additional complexity related to the distance between consecutive stimuli and the capacity of the system between stimuli to pass from the transient to the steady-state. {The Flip Flop is not only a binary decision, and the distribution of eigenvalues in the complex plane frequently has more of them outside the unit circle compared to the other tasks.}

{In most situations, it is observed} that a fixed point state corresponding to the real eigenvalue appears and {also} a complex conjugate pair, which is generally related to the passivated state of the output.

\begin{figure*}[htb!]
\begin{center}

\includegraphics[width=12.5cm]{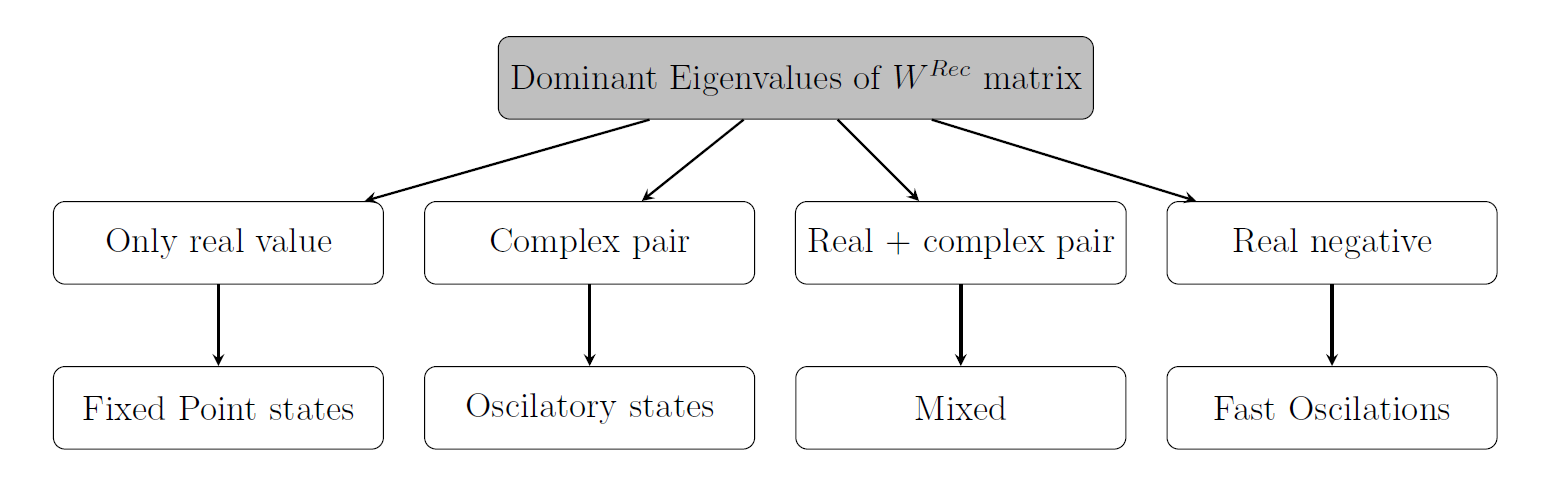}
\caption{ {Summary of the behaviour observed for the trained network's activity of the units, $h_i(t)$, when receiving different stimuli linked with the dominant eigenvalues outside the unit circle. These results are summarized considering the four possible tasks for both initial conditions before training. Examples of each behavior are avaliable in the \textbf{Supplementary Information}.}\label{fig_10}}
\end{center}
\end{figure*}

Considering the {tasks studied and the different realizations obtained}, in general, Figure \ref{fig_10} {shows} a schema that allows us to summarize the results, linking the dominant eigenvalues with the behaviour of the activity.

{According to the results presented in this section, } {there are multiple ways for the internal configurations to reproduce the same task for which the network was trained. In each task, it is necessary to have at least one mode, of those defined before, associated with each possible output state. Additional modes may exist, but these will not be excited by the input stimulus combinations.}

{Following the description presented in Section \ref{long}, and in the light of the results obtained in the simulations shown in Section \ref{RESULTADOS NUMERICOS}, the states described in the Equations \ref{eq-lin-4} and \ref{eq-lin-5} were observed. In fact, for the realizations that have complex dominant eigenvalues, if we  numerically estimate the frequency of oscillation, of  the  activity $ h_i(t) $, it is approximately:}

{
\begin{equation}
f= \frac{1}{2 \pi} \frac{ Im(\lambda _{max})} { Re ( \lambda_{max})}
\label{eq-lin-6}
\end{equation}
}

Which is consistent with estimates made in \cite {10.1371/journal.pcbi.1006309, PhysRevE.88.042824}.

\subsection{ { More about dynamics inferred from $\mathbf{W^{Rec}}$ }}\label{linking}

{In \cite{Asllanieaau9403} authors studied dynamical processes evolving on non-normal networks and have shown how small disturbances may undergo a transient phase and be strongly amplified in linearly stable systems. Such property (not normal matrices) was not selected nor expected to be necessary for solving the particular tasks studied here. But in all realizations obtained of present work (also the realizations that started initially with orthogonal or normal matrices), at least a minimum degree of non-normality was found after the training.}

The matrices of a trained networks are not normal, as shown in eigenvalue distributions of the examples in Figure \ref{fig_09}. The previous analysis is not fully complete. Although the matrices of the simulations are approximately normal when considering orthogonal condition (see Appendix \ref{B}) since they do not deviate much from the initial condition after training. They are enough not-normal so that there is a transient amplified effect that leads the system from the initial condition to the long term dynamics observed. This happens for all realizations (see Appendix \ref{B} for more details). 

The departure from the normal condition of the matrix can be estimated through the parameter Henrici's departure from normality, obtained as in Equation \eqref{eq-lin-7}.

\begin{equation}
d_F(\mathbf{W^{Rec}})=\frac{\sqrt{(||\mathbf{W^{Rec}}||^2-\sum^N_{i=1}|\lambda_i|^2)}}{||\mathbf{W^{Rec}}||}
\label{eq-lin-7}
\end{equation}

Where, for normalization, it is divided by the norm of the matrix.

The long term dynamics was previously obtained through linearization. The departure from normality is which leads the system from equilibrium to the final state and makes appear more complex patterns for the activity \cite{Asllanieaau9403}.

It was observed, in some realizations, that appear high-frequency oscillations that sometimes include modulations.

{Following \cite{10.1371/journal.pcbi.1007655}, the observed transient in the activity can be also related to the norm of $ \mathbf{h(t)} $. This norm is the euclidean distance between the equilibrium point of the system and the activity at time $ t $.} It is estimated as:

\begin{equation}
||\mathbf{h(t)}||=\sqrt{\sum^N_{i=1} \tilde{h}^2_i(t)+2\sum^N_{i>j}\tilde{h}_i(t)\tilde{h}_j(t)\mathbf{v_i}\mathbf{v_j}}
\label{eq-lin-8}
\end{equation}

This magnitude has been previously studied as an amplification mechanism in neural signals \cite{10.1371/journal.pcbi.1007655}, where authors studied the change or the slope of the $ \mathbf{h (t)}$ norm, and the conditions for the appearance of amplified trajectories like the ones observed in the present work. They affirm that the necessary condition for having amplified trajectories is on the eigenvalues of the symmetric part of the matrix $\mathbf {W^{Rec}} $ estimated as in \eqref{eq-lin-9}. This condition establishes that the maximum eigenvalue of the symmetric part of the matrix must be greater than 1.

\begin{equation}
 \mathbf{W^{Rec}_{sym}}=\frac{1}{2}(\mathbf{W^{Rec}+{W^{Rec}}^T})
\label{eq-lin-9}
\end{equation}

{Let us remember that symmetric matrices have all their real eigenvalues. For all the realizations in the simulations of present work, the maximum eigenvalue of the symmetric matrix is always greater than 1, therefore the condition for the existence of transients is guaranteed}. Only some specific values of the initial condition, $ h_ {o, i} $, will be amplified according to \cite{10.1371/journal.pcbi.1007655}. This is consistent with the observations that when networks are elicitated with different amplitude values for the input pulse there is an amplitude limit for which the paths are not amplified anymore.

In the case of the realizations obtained, a transient ending in a sustained oscillation, or one going to a fixed point different from zero is always observed. Exceptionally for tasks with a passivated state for the output attenuation is observed.

In general, the behavior of the system when eigenvalues are lying outside the unit circle, either with the real part less than 1 or with the negative real part, is to present rapid oscillations. In those cases, the system seems to be also governed by the set of eigenvalues outside the unit circle since the modes that remain within tend to attenuate the transients.

{In an RNN, an eigenvalue outside the unit circle could give a chaotic attractor state, but it is not the case of the trained networks presented here (with all parameters trained) because a chaotic attractor will not be able to provide stationary output needed according to training rule.}

{Regarding the spectral radius, it is not possible to train all the network's parameters with few epochs starting with weight distributions with an initial value of spectral radius $g$ much larger than 1, or even much smaller. 
To generate networks with chaotic behaviour, but still capable of fulfilling the tasks, another paradigm is possible such as reservoir computing, where only the output weights are trained. 
In \cite{Laje2013} authors considered networks with chaotic behaviour but where only the output parameters are trained. The number of units must be larger for the reservoir. In this case other training methods are used, and then different values for $g$ can be considered to include chaotic behaviour.}

\section{Discussion \label{DISCUSION}}

The brain represents sensory stimuli with the collective activity of thousands of neurons. Coding in this high-dimensional space is typically examined by combining linear decoding and dimensionality-reduction techniques \cite{Cunningham2014, Bagur2018, 10.1371/journal.pcbi.1007655}, as it was explored in the current paper. The underlying network is often described in terms of a dynamical system \cite{SUSSILLO2014156, BARAK20171, Mante2013}. Here a set of task-based trained networks was analyzed, which has become a popular way to infer computation functions of different parts of the brain \cite{NIPS20199694}. The aim was to classify the obtained behaviors and relate {them} to the eigenvalue spectrum.

Without any constraint, different behaviors for the same task were obtained. Some were observed already in different experimental data sets. That includes oscillatory activity \cite{NAMBU20201} and fixed points memories in the space state \cite{DBLP:journals/neco/SussilloB13}.

The results presented suggest that it is not sufficient to implement a task-based approach only, without considering other aspects related to the brain system or part that we want to model, given that multiple dynamical systems could represent the same temporal task. On the contrary, to do this, it is necessary to consider additional information to constraint the model and be able to compare it thoroughly \cite{10.1007/978-3-030-61609-0_69}.

{Let us consider the results of \cite{NIPS2019_9694}. In this work, the authors studied networks trained to perform three standard neuroscientific-inspired tasks and characterized their non-linear dynamics.} 

{On the one hand, they found that the geometry of the representation of the RNN is highly sensitive to the choice of the different architectures (RNN, GRU or LSTM), which is expected since the equations of these architectures, and the internal states, are different from each other. For this study, they used different measures of similarity, such as Canonical Correlation Analysis \cite{NIPS2019_9694}.}

{On the other hand, they found that, while the geometry of the network can vary throughout the architectures, the topological structure of the fixed points, transitions between them, and the linearized dynamics appear universally in all architectures.} 

{However, in \cite{NIPS2019_9694}, it is not stated that each topological structure is linked uniquely to each task. The results of the present work do not contradict those of [4], they extend the analysis showing how, within the same network architecture (in this case RNN), different topological structures can be obtained, as long as at least one structure (and one transition) is associated with the different decisions for which the network has been trained to take in the task.}

{For example, for the case of the 3 bits Flip Flop task, that was studied in \cite{NIPS2019_9694}, we have obtained preliminary results \cite{jarne2021need}, where we also observed a topology similar to that of Figure 1 of \cite{NIPS2019_9694}, with eight fixed points. However, we have also obtained different spatial distributions and orientations of these points for the different realizations.}

{In simpler tasks related to binary decisions, such as the tasks studied in this work, the system has more freedom to achieve the same result through training. The minimum necessary condition for the system is that the recurrent weight matrix has one real eigenvalue (or a conjugate pair of complex eigenvalues) outside the unit circle associated with each state that the system must take to fulfil the task. In that sense, there is universality, but not in the sense that the number of eigenvalues or their value associated with each state within the task is unique.}

{In the way that the network is configured due to training, the activation function (hyperbolic tangent) allows the activity $h_i(t)$ of units to grow and saturate to a value that converges to a fixed amplitude (fixed-point), or ultimately allows small oscillations. In both cases, $\mathbf{W^{out}}$ combines them finally to decide} {according to} {the stimuli for which the network was trained. These options are reflected in the eigenvalue spectrum with the values outside the unit circle.}

{Because it is a task-based approach, in principle, for the trained networks, nothing restricts the different possible internal configurations that give rise to the same decision-making process. This is true for both initial configurations considered here: Normal Random and Orthogonal Random. The diversity of solutions does not depend on the training algorithm (backpropagation through time with ADAM optimizer was used in present work as shown in Section \ref{MODELO}). Different configurations also arise for similar tasks using other training techniques such as Force and Full-force for example} {used} {in \cite{DBLP:journals/neco/SussilloB13, 10.1371/journal.pone.0191527}.}

{However, regularization methods could be used to penalize some of the solution types against others if one had some argument or hypothesis related to the dynamics of the biological process that motivates it. In some studies, RNNs are optimized to reproduce neural data. Other studies take a task-oriented approach, such as present work, by training a network to perform a task and then attempting to find similarities between the RNN’s population dynamics and those of biological neurons recorded from an animal performing a similar task. It is also possible to take a hybrid approach, by training a network on a task with constraints that yield more brain-like solutions \cite{Pollock2019.12.19.883207}.}

{Regarding the solutions obtained in the tasks considered in this work, which form part of the set of temporary decision-making tasks, increasing the number of units does not favor any kind of solution over another \cite{jarne2019detailed}. Also, there are no differences in terms of accuracy if we compare the fixed point solutions with the oscillatory ones.}

{In more complex tasks where, for example, we can consider multi-tasking as more complex \cite{Jarne_2021}, what happens is that it is necessary to add more units for the network training to converge.}

\section{Conclusions \label{CONCLUSIONS} }

Considering the analysis made above, it is possible to highlight some aspects of the results obtained in the study. On one hand, networks trained for these four tasks (AND, XOR, OR, and Flip Flop) have consistent patterns, and they are not stable systems, { in the sense described in \cite{10.1371/journal.pcbi.1007655}, meaning that the trajectory asymptotically decays to the equilibrium state that corresponds to $h_i(t=0)= 0$}, which in principle is not an unexpected situation. The classification for the set of tasks proposed here and its dynamics are interesting since these tasks could constitute possible flow control mechanisms for information in the cortex.

{On the other hand, different realizations for the same task were obtained with different dynamical behaviours, and the trained networks are generally non-normal \cite{sengupta2018robust}.}

{It is interesting to note that as indicated in Section \ref{DISCUSION}, a priori, the obtained dynamics was not created on purpose. It was not created by using input-driven attractor states, either fixed points or limit cycles, and finding appropriate projection from the obtained attractor states to required output state learning $\mathbf{W^{out}}$.  Rather the dynamics of the system arises as a result of requesting the system to learn the task, with the considered initial conditions.}

Linearization was a useful mechanism to understand the behaviour of the system in the first order. Thus the decomposition into eigenvalues of the matrix of recurrent weights is linked with the activity for these tasks.

The results obtained support the hypothesis that trained network represents the information of the tasks in a low-dimensional dynamics implemented in a high dimensional network or structure \cite{BARAK20171} as also reported in \cite{Kuroki2018}. 

The neural network model studied in this work, as described in Section \ref{MODELO}, is widely used to describe results in different experiments and areas in neuroscience. For example in motor control \cite{doi:10.1152/jn.00467.2018}. In particular, analyzes on the cerebral cortex show complex temporal dynamics \cite{SUSSILLO2014156, DBLP:journals/neco/SussilloB13, nature_01, nature_com}, where different mechanisms to control the information flow could be present and coexist. For this reason, knowing the details of the model's dynamics is important to understand the observed experimental results with greater precision.

Future extensions of the present work will include the distinction between excitatory and inhibitory neurons.

\appendix

\section{Appendix: Variation of the distribution's moments \label{A}}

Table \ref{tabla_fit_01} and \ref{tabla_fit_02} show the changes in the moments of distribution after training for the realizations of each task. It is not possible to estimate the differences of variance with a fit due to the variations of less than 0.1\% between initial condition and trained networks (points are too close to perform a meaningful fit). 

\begin{table*}[htb!]
\begin{center}
\begin{small}
\begin{tabular}{lllllllll}
\hline
\textbf{Moment}&\textbf{And} & \textbf{Xor} & \textbf{Or} & \textbf{FF} & \textbf{$\Delta$ And} & \textbf{$\Delta$  Xor} &\textbf{ $\Delta$ Or} & \textbf{ $\Delta$ FF} \\ \hline
$\mu$&-0,9 & -0,009 & 0,83  & -4,8 & 0,02 & 0,01 & 0,002 & 0,03 \\ \hline
Variance &- & -      & -  & -    & -    & -    & -     & -    \\ \hline
Skewness& 14  & 4  & -0,95 & 2,32 & 0,7  & 1    & 0,2   & 0,9  \\ \hline
Kurtosis &15  & 0,82  & 1,47   & 3,85 & 0,7  & 1    & 0,2   & 0,98 \\ \hline
\end{tabular}
\end{small}
\caption{Percentage variation for each moment and task, {with} its uncertainties with respect to the initial condition Orthogonal pre-training.}
\label{tabla_fit_01}
\end{center}
\end{table*}

\begin{table*}[htb!]
\begin{center}
\begin{small}
\begin{tabular}{lllllllll}
\hline
\textbf{Moment}&\textbf{And} & \textbf{Xor} & \textbf{Or} & \textbf{FF} & \textbf{$\Delta$ And} & \textbf{$\Delta$ Xor} & \textbf{$\Delta$ Or} & \textbf{$\Delta$ FF} \\ \hline
$\mu$&-5,82   & -2,95  & -0,85 & -2  & 0,01 & 0,01 & 0,005 & 0,01 \\ \hline
Variance &-   & -   & -    & -     & -    & -    & -     & -  \\ \hline
Skewness &6,48   & 0,37  & 10,9  & 9,96 & 1,2  & 0,11 & 1     & 1  \\ \hline
Kurtosis &4,40 & -0,22 & 12,26   & 2,46    & 1    & 0,18 & 1,03  & 0,6  \\ \hline
\end{tabular}
\end{small}
\caption{Percentage variation for each moment and task, {with} its uncertainties with respect to the initial condition Random Normal pre-training.}
\label{tabla_fit_02}
\end{center}
\end{table*}

In each cell of the table is included the moment for each tasks. Results are obtained with linear regression where the x-axis is the initial value, and the y-axis is the value after trained is performed. The departure from the identity line is measured in percentage with its uncertainty $\Delta$. The positive mean value is larger with respect to the initial condition and the negative means smaller. Each cell of the table represents the fit result of the moment, considering the set initial-final of all realizations.

\section{Appendix: Henrici's number \label{B}}

\begin{figure}[htb!]
\begin{center}

\includegraphics[width=7cm]{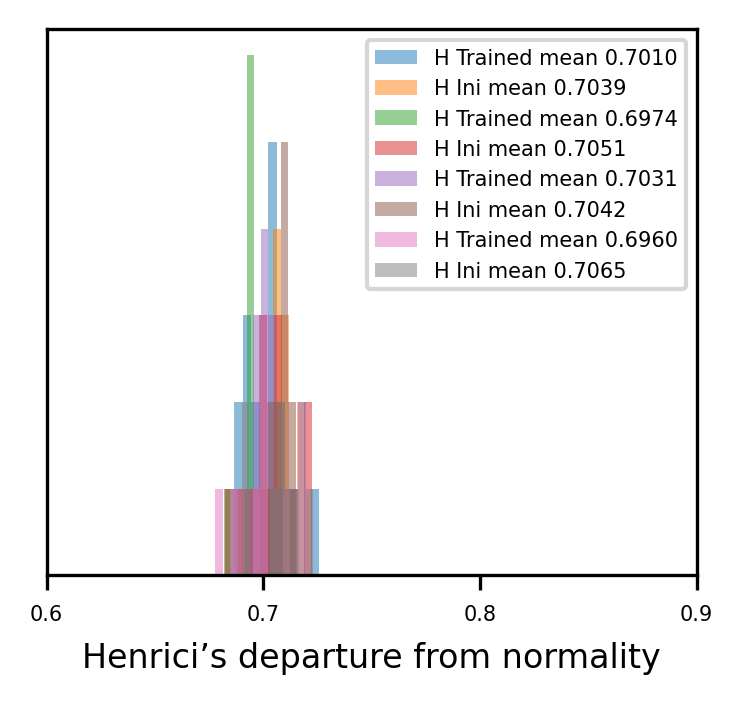}
\includegraphics[width=7cm]{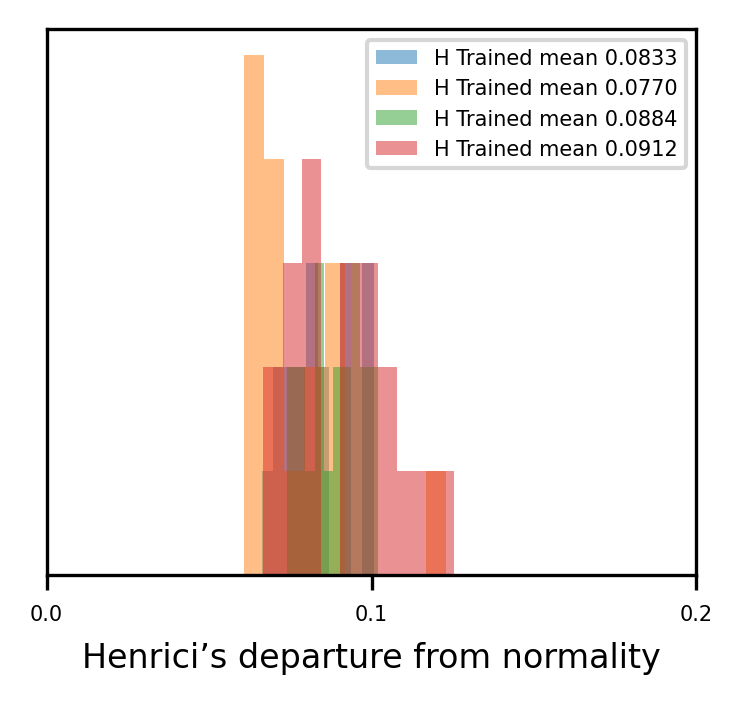}
\caption{Histograms with Henrici's number for each task (different colors represent the four tasks). The figure on the bottom corresponds to the Random Normal condition. Also, values for the initial condition are included. The figure on the top corresponds to the Orthogonal condition. \label{fig_11}}
\end{center}
\end{figure}

The histograms of Figure \ref{fig_11} show, separating the tasks {by colour}, the averages of Henrici's numbers calculated for the matrices of each of the tasks (AND, OR, XOR, and Flip Flop). The values on the bottom of Figure \ref{fig_11} correspond to the matrices trained from the orthogonal condition, and those from the top correspond to the random normal condition. It is observed that, when the initial condition is the same, the values obtained in the different tasks do not present significant differences between them.

\section{Supplementary Information}\label{sup-c}

Code, simulations, and additional figures of this analysis are available at the following Github repository: \\

\url{https://github.com/katejarne/RRN\_dynamics}\\

It is a public repository. Also, the Statistical Summary (summary.pdf) includes additional examples and guidelines to navigate the repository content that includes simulation files and details.

\vspace{0.5cm}

\textbf{Conflicts of interest.} The author declare that she has no conflict of interest.

\section*{Acknowledgments}

Present work was supported by CONICET and UNQ. This research did not receive any specific grant from funding agencies in the public, commercial, or not-for-profit sectors. I thanks mainly the researchers who encourage me to keep working in neural networks models. {In particular, many thanks to Mariano Caruso for shear with me valuable discussions.} {I also thank the anonymous reviewers for their careful reading of the manuscript and their insightful comments and suggestions.} Finally, I want to thanks Gabriel Lio, whose support during the Covid-19 quarantine was fundamental.

\Urlmuskip=0mu plus 1mu\relax
\bibliographystyle{spphys} 
\bibliography{mybibfile}

\end{document}